\documentclass{emulateapj-rtx4}
\shorttitle{On the challenging variability of LS IV-14$^\circ$116}
\shortauthors{Miller Bertolami M. M., C\'orsico A. H. \& Althaus L. G.}
\begin{document}
\title{On the challenging variability of LS IV-14$^\circ$116:\\
 pulsational instabilities excited by the $\epsilon$-mechanism} 
\author{M. M. Miller Bertolami$^{1,2}$, A. H. C\'orsico$^{1,2}$ 
\& L. G. Althaus$^{1,2}$} 
\affil{  $^{1}$Facultad de Ciencias Astron\'omicas y Geof\'isicas,
  Universidad Nacional de La Plata, Paseo del Bosque s/n, 1900 La
  Plata, Argentina.\\ 
$^{2}$Instituto de Astrof\'isica La Plata
  (CCT-La Plata, UNLP-CONICET), Paseo del Bosque s/n, 1900 La Plata,
  Argentina.}  
\email{mmiller@fcaglp.unlp.edu.ar}

\begin{abstract}

We investigate the pulsation driving mechanism responsible for the
long-period photometric variations observed in LS IV-14$^\circ$116, a
subdwarf B star showing a He-enriched atmospheric composition.  To
this end, we perform detailed nonadiabatic pulsation computations over
fully evolutionary post He-core-flash stellar structure models,
appropriate for hot subdwarf stars at evolutionary phases previous to
the He-core burning stage.  We found that the variability of LS
IV-14$^\circ$116 can be attributed to nonradial $g$-mode pulsations
excited by the $\epsilon$-mechanism acting in the He-burning shells
that appear before the star settles on the He-core burning stage. Even
more interestingly, our results show that LS IV-14$^\circ$116 could be
the first known pulsating star in which the $\epsilon$-mechanism of
mode excitation is operating. Last but not least, we find that the
period range of destabilized modes is sensitive to the exact location
of the burning shell, something that might help to distinguish between
the different evolutionary scenarios proposed for the formation of
this star.

\end{abstract}


\section{Introduction}
About 5\% of all hot subdwarf stars (sdB, sdO) show He-enriched
surface abundances (He-sdB, He-sdO). While most normal H-rich
subdwarfs are supposed to be low mass core He-burning stars with
atmospheres dominated by H due to the action of gravitational
settling, the evolutionary status of the He-rich subclass is less
clear. He-rich subdwarfs have been suggested to be the result of
either He white dwarf mergers \citep{2000MNRAS.313..671S} or late
helium core flashes \citep{2001ApJ...562..368B}. Regardless the
particular scenario proposed for their formation, it is accepted that
some of these stars are still contracting towards the He-core burning
phase (EHB), as otherwise gravitational settling of the remaining H
would have turned the star into a H-rich sdB star.

Two families of pulsators have been discovered within the H-rich sdB
stars: the rapid pulsators (sdBVr; \citealt{2010IBVS.5927....1K})
discovered by \cite{1997MNRAS.285..640K}, and the slow pulsators
(sdBVs; \citealt{2010IBVS.5927....1K}) discovered by
\cite{2003ApJ...583L..31G}. While sdBVr stars show short pulsation
periods ($\sim 80-400$ s) ascribed to radial modes and non-radial
$p$-modes, sdBVs pulsations (with periods $\sim 2500-7000$ s) are
associated to non-radial long period $g$-modes. Pulsations in both
groups of variable stars have been explained by the action of the
$\kappa$-mechanism due to the partial ionization of iron group
elements in the outer layers, where these elements are enhanced by the
action of radiative levitation \citep{1997ApJ...483L.123C,
  2003ApJ...597..518F}.  While many sdBV stars are known, only one
He-sdB star, LS IV-14$^\circ$116, has been found to be a pulsator
\citep{2004Ap&SS.291..435A, 2005A&A...437L..51A}. In fact, LS
IV-14$^\circ$116 is a very intriguing object: while its atmospheric
parameters ($T_{\rm eff}$ and $g$) place it well within the sdBVr
instability region \citep{2010MNRAS.409..582N, 2011ApJ...734...59G},
it displays periods typical of the sdBVs family of pulsators
\citep{2005A&A...437L..51A, 2011ApJ...734...59G}. From a
spectroscopical perspective, LS IV-14$^\circ$116 is also an intriguing
object, showing a mild He-enrichment ($n_{\rm He}=0.16$) and very high
excesses of zirconium, yttrium and strontium in its atmosphere
\citep{2011MNRAS.412..363N}. In this sense, both the He-enrichment and
the heavy metal excesses might be related to the effects of ongoing
diffusion before reaching diffusive equilibrium, thus placing the star
in the pre-EHB phase.  Up to now, the driving mechanism behind the
long period pulsations of LS IV-14$^\circ$116, as well as the absence
of short period pulsations, remains a mistery and the star poses a
challenge to the theory of stellar pulsations
\citep{2008ASPC..392..231F}.  In this connection, the recent
confirmation of both the multiperiodic variability and the $T_{\rm
  eff}-g$ values for LS IV-14$^\circ$116 \citep{2011ApJ...734...59G}
strongly increase the enigma.

While most pulsations exhibited by pulsating stars (included those of
sdBVs and sdBVr stars) are self excited through the classical
$\kappa$-mechanism, no star has been discovered so far to be excited
by the $\epsilon$-mechanism. The only possible exception are some
oscillations in $\delta$ Scuti stars \citep{2006CoAst.147...93L} and
PG 1159 stars \citep{1986ApJ...306L..41K, 2009ApJ...701.1008C}, but
neither of these have been confirmed. In the $\epsilon$-mechanism, the
driving is due to the strong dependence of nuclear burning rates on
temperature. In the layers where nuclear reactions take place, thermal
energy is gained at compression by the enhancement of nuclear energy
liberation, while the opposite happens during the expansion phase
\citep{1989nos..book.....U}.

In the present letter we show that long period $g$-mode pulsations,
like those observed in LS IV-14$^\circ$116, can be understood as
driven by the $\epsilon$-mechanism acting during the unstable shell
burning events that take place before the star settles on the He-core
burning phase.  In addition, our scenario might explain the intriguing
absence of short period $p$-mode pulsations in LS IV-14$^\circ$116. In
the light of our results, LS IV-14$^\circ$116 would be the first star
in which pulsations driven by the $\epsilon$-mechanism have been
detected, thus being the first proof that the $\epsilon$-mechanism can
indeed drive pulsations in stars.

\section{Scenario and numerical details}

\begin{table}
\begin{center}
\begin{tabular}{ c| c c c }
\# subflash & $\tau_{\rm duration}$ &  shortest $\tau_{\rm e}$ & $\tau_{\rm glo}$ \\
            & [yr]                &  [yr]        & [s]               \\\hline
1     &  $\sim 2015  $&$       110 $&    31856     \\
2     &  $\sim 4674  $&$       250 $&    33928     \\
3     &  $\sim 7978  $&$       610 $&    46402     \\
4     &  $\sim 11923 $&$      1430 $&    63176     \\
5     &  $\sim 18286 $&$      1970 $&    85172     \\
6     &  $\sim 23300 $&$      2000 $&   129910     \\
7     &      -        & -           &   241674     \\
\end{tabular}
\caption{Tipical timescales during the He-subflashes before the
  begining of the EHB.  $\tau_{\rm duration}$ indicates the length
  of the unstable stages during each He-subflash, as shown in Fig.
  \ref{Fig:Peri_Lumi}. $\tau_{\rm e}$ is the e-folding time obtained
  from the nonadiabatic pulsation computations.  $\tau_{\rm glo}$ is
  a global estimation of the convective turnover time (see text).}
\label{Tabla}
\end{center}
\end{table}

To explore the effects of the $\epsilon$-mechanism in a contracting
pre He-core burning subdwarf star, we have computed the evolution of
an initially 1.03 $M_\odot$, $Z=0.02$ star model (see
\citealt{2008A&A...491..253M} for details about this sequence) that
departs from the red giant branch just before the He-core flash, and
experiences the flash as a shallow mixing hot flasher of 0.47378
$M_\odot$ (mild He-enrichment, $Y=0.5978$; see
\citealt{2004ApJ...602..342L} for a definition). Stellar structure and
evolution equations have been solved with {\tt LPCODE} stellar
evolution code as in our previous works (\citealt{2008A&A...491..253M}
and references therein). Although our computations have been performed
within the hot-flasher scenario, the existence of He-shell subflashes
in both the hot-flasher and merger scenarios makes our approach
valid regardless the actual origin of LS IV-14$^\circ$116.

\begin{figure}
\begin{center}
\includegraphics[clip, angle=0, width=7.5cm]{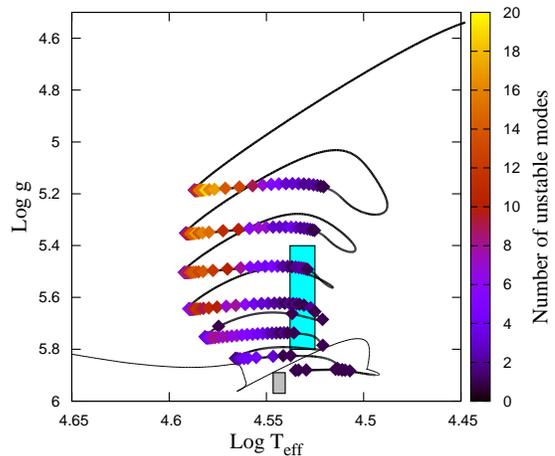}
\caption{Evolution of our sequence before settling on the He-core
  burning phase. Thick line indicates pre-EHB evolution while thin
  line corresponds to the evolution after the EHB. Colored points indicate
  the location of models for which unstable modes have been
  found. These unstable modes are caused by the $\epsilon$-mechanism
  acting in the unstable He-burning subflashes that follow the primary
  He-core flash.  Color coding indicates the number of unstable modes
  found in each model. Boxes indicate the location of LS
  IV-14$^\circ$116 according to \cite{2010MNRAS.409..582N} (cyan) and
  \citep{2011ApJ...734...59G} (grey).}
\label{Fig:Teff-g} 
\end{center}
\end{figure}

As seen in Fig. \ref{Fig:Teff-g}, after the primary He-core flash (not
shown in the figures) the star contracts towards the He-core burning
phase through a series of He-shell subflashes (loops in
Fig. \ref{Fig:Teff-g}), evolving through the region in the $T_{\rm
  eff}-g$ diagram occupied by LS IV-14$^\circ$116. For this sequence
we analyzed the pulsational stability properties of more than 1000
stellar structure models, one every five stellar evolution timesteps,
covering the whole evolution from the primary He-core flash to the final
settling on the He-core burning phase. For each model we restricted
ourselves to explore the $\ell=1$ $g$-modes in the period range from
500 to 6000 s. The stability analysis of the stellar models was
carried out by means of the linear, nonradial, nonadiabatic pulsation
code described in \cite{2006A&A...458..259C} with the inclusion of the
$\epsilon$-mechanism mode driving operating in He-burning layers as
described in \cite{2009ApJ...701.1008C}.
 Two standard simplifications have been introduced. We have adopted
 the ``frozen-in convection'' approximation in the stability analysis
 and we have assumed that $dS/dt=0$ in the non-perturbed background
 model adopted in the computations. While the latter is not rigorously
 true in contracting stellar models, the timescale of the
 He-subflashes ($\sim 1000-10000$ yr) is nonetheless much longer than
 the oscilation periods ($500-6000$ s), and this approximation is
 correct. To assess the validity of ``frozen-in convection''
 approximation we computed the turn-over timescale of convection at
 the peak of the helium flashes as
\begin{equation}
\tau_{\rm glo}=\int_{R_{\rm base}}^{R_{\rm top}}  \frac{dr}{v_{\rm MLT}},
\end{equation}
where $R_{\rm base}$ and $R_{\rm top}$ are the radial coordinates of
the boundaries of the convective zone and $v_{\rm MLT}$ is the
velocity of the convective elements within the mixing length theory of
convection. As seen in the fourth column of Table \ref{Tabla}, the
convective turnover times are much longer than the unstable periods
obtained in this work (see Fig. \ref{Fig:Peri_Lumi}). Thus, the
assumption of ``frozen-in convection'' is largely justified.

\section{Results}
\begin{figure}
\begin{center}

\includegraphics[clip, angle=0, width=7.2cm]{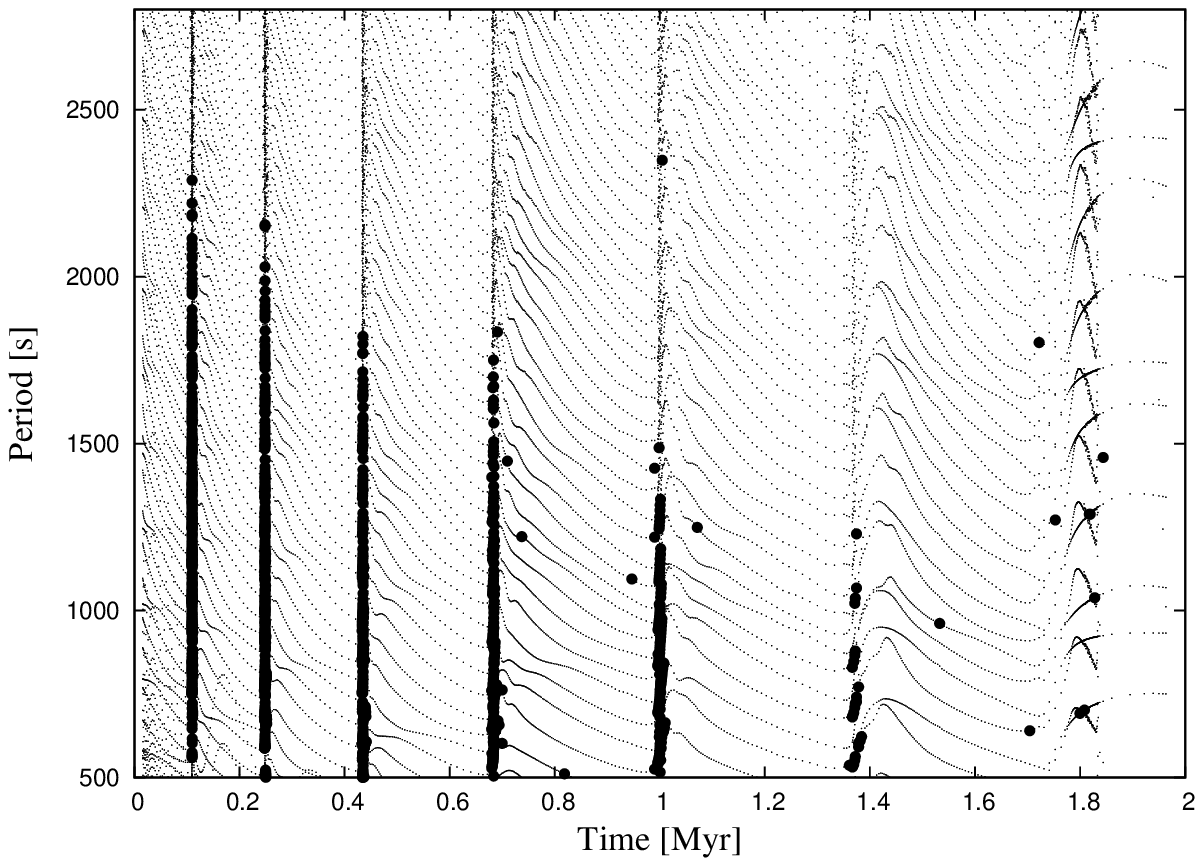}\\

\includegraphics[clip, angle=0, width=7.5cm]{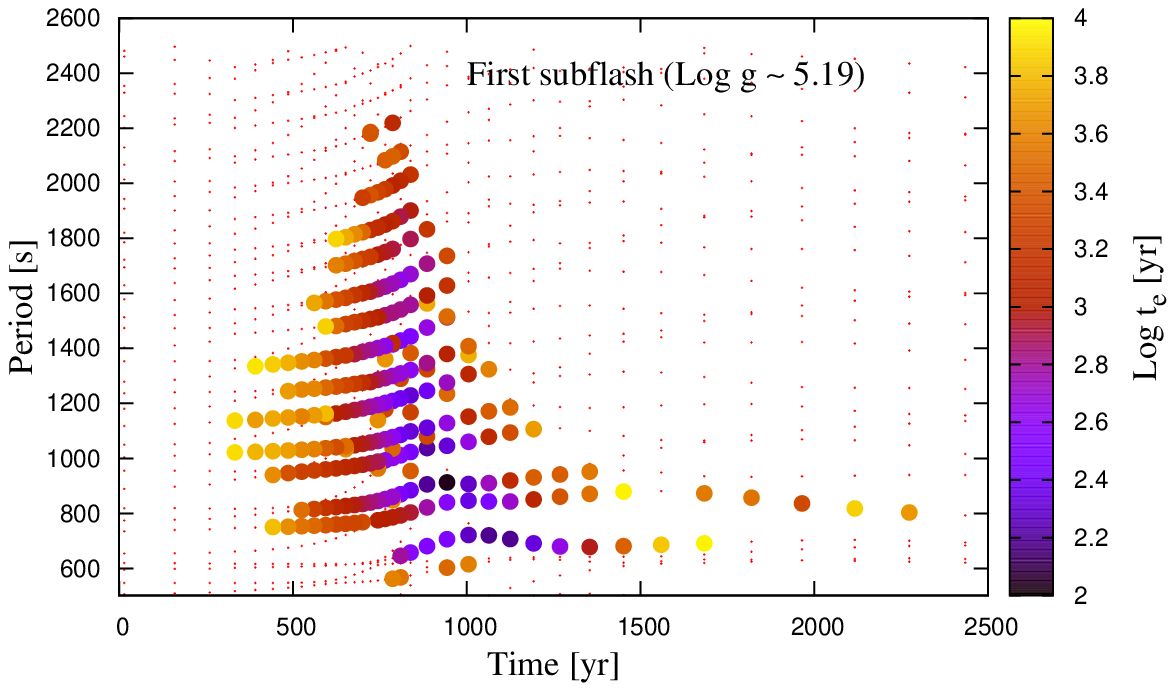}\\

\caption{{\it Upper Panel:} Temporal evolution of the pulsation periods of our
  He-sdB models during the He-subflashes before the EHB. Thick
  points indicate those periods found to be formally unstable in our
  computations. {\it Lower Panel:} The situation for the first and most
  intense subflash. Colored points indicate unstable modes with color
  coding indicating the e-folding time of each unstable mode.  }
\label{Fig:Peri_Lumi} 
\end{center}
\end{figure}
\begin{figure}
\begin{center}
\includegraphics[clip, angle=0, width=7.5cm]{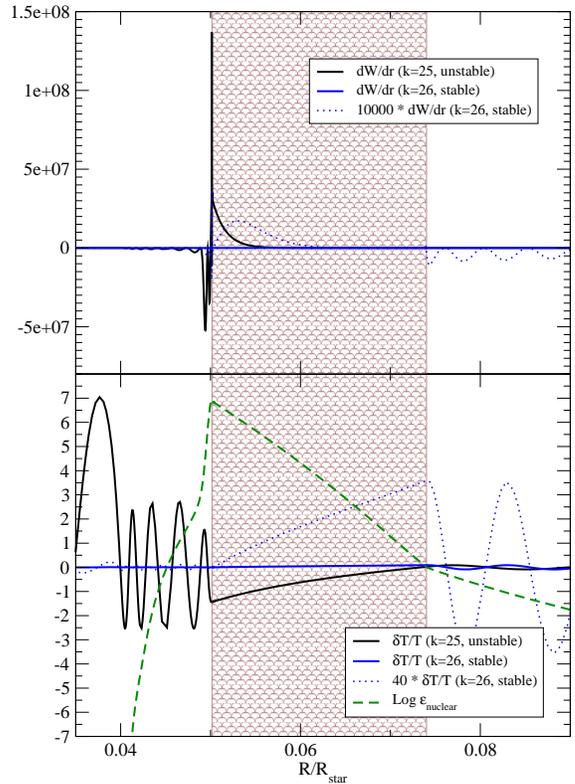}
\caption{ {\it Lower Panel:} Values of $\delta T/T$ for two
  consecutive modes during the first He-subflash, $k=25$ (unstable)
  and $k=26$ (stable). For the unstable mode, the large amplitude of
  $\delta T/T$ in the region of the He-burning shell can be easily
  appreciated.  {\it Upper Panel:} Differential work ($dW/dr$) for the
  modes $k=25$ and $k=26$. Note the strong driving contribution of the
  burning shell ($dW/dr>0$) in the case of the unstable mode.  In the
  interest of comparison we have arbitrarily magnified $\delta T/T$
  and $dW/dr$ for the stable mode (dotted line).  Brown curls indicate
  the location of the He-flash driven convective zone.}
\label{Fig:modos} 
\end{center}
\end{figure}

In Fig. \ref{Fig:Teff-g} we can see how the star becames recurrently a
pulsator as it evolves towards the EHB. Our sequence of models is very
close to the location of LS IV-14$^\circ$116 in the $T_{\rm eff}-g$
diagram. Note that no attempt to fit the exact location has been done,
as we have just used a sequence from our previous work
\citep{2008A&A...491..253M}. It would be easy to attain a better
agreement by considering a sequence with a lower He-abundance or
earlier He-core flash (see \citealt{2008A&A...491..253M}).
Fig. \ref{Fig:Peri_Lumi} shows the evolution of the pulsation periods
of our He-sdB sequence of models. In the upper panel we depict the
situation found for the complete pre-EHB phase. Note that, each time a
subflash develops, we obtain unstable $g$-modes, with the periods
getting shorter as the star experiences subsequent, deeper,
subflashes. The bottom panel illustrates the situation for the first
and strongest flash of the series and shows the derived e-folding
times for each mode. Notably, modes are continuously destabilized
during the whole flash, with e-folding times ($\tau_e$) of the excited
modes getting shorter than the duration of the pulse ($\tau_{\rm
  duration}$), indicating that these modes should attain observable
amplitudes (see also Table \ref{Tabla}).

While the whole evolution from the primary He-flash to the EHB takes
about 2 Myr, the unstable stages add up to $\sim 68000$ yr (see Table
\ref{Tabla}). It is worth noting that this refers to the formally
unstable stage, including some stages with somewhat long e-folding
times, and that the actual unstable phases might be shorter.  Then,
within this picture, we would expect the variable to non variable
ratio of He-sdB to be lower than $1/30$. It is worth noting that this
is consistent with the fact that only one He-sdB star, out of 23
studied objects, has been found to be a variable
\citep{2005A&A...437L..51A, 2004Ap&SS.291..435A}.  As shown in Fig.
\ref{Fig:Peri_Lumi}, most unstable modes exhibit periods in the range
$600-2000$ s close to the periods observed in LS IV-14$^\circ$116
(1954-5084 s), and definitely much longer than the periods displayed
by sdBVr stars in the same region of the $T_{\rm eff}-g$ plane. In
particular, it is worth noting that our models are able to reproduce
the dominant periodicity observed in LS IV-14$^\circ$116 ($P=1954$ s).

Our computations show that not all modes with periods within the
unstable range are excited. In fact, only modes with significant
amplitudes in the He-burning shell are destabilized. This can be seen
in Fig. \ref{Fig:modos} where the real part of the eigenfunction
$\delta T/T$ and the differential work ($dW/dr$) for two modes with
consecutive radial orders ($k=26$; stable and $k=25$; unstable) are
shown. This behavior, in which the $\epsilon$-mechanism acts as a
narrow band filter exciting only modes with large amplitudes in the
narrow He-burning shell region, has been reported in previous works
\citep{1986ApJ...306L..41K, 2009ApJ...701.1008C}.

It has been mentioned \citep{2011ApJ...734...59G} that the effect of
the enhancement of He in nonadiabatic computations in the context of
the $\kappa$-mechanism in sdB stars might be related with the
pulsations observed in LS IV-14$^\circ$116. As a consequence, once
might wonder to which extent our results are related with the
$\kappa$-mechanism. Also, one might wonder to which extent the so
called ``convective blocking'' mechanism \citep{1987ApJ...314..598P}
at the He-shell flash convective zone might help to destabilize the
modes.  In order to check our results, we have carried out explicit
tests by switching off the derivatives of $\epsilon_{\rm He}$ in our
nonadiabatic computations. As expected, when the $\epsilon$-mechanism
is turned off, all unstable modes shown in Fig. \ref{Fig:Peri_Lumi} become
stable, indicating that the driving mechanism behind the unstable
modes is indeed the $\epsilon$-mechanism alone.

\begin{figure}
\begin{center}
\includegraphics[clip, angle=0, width=7.5cm]{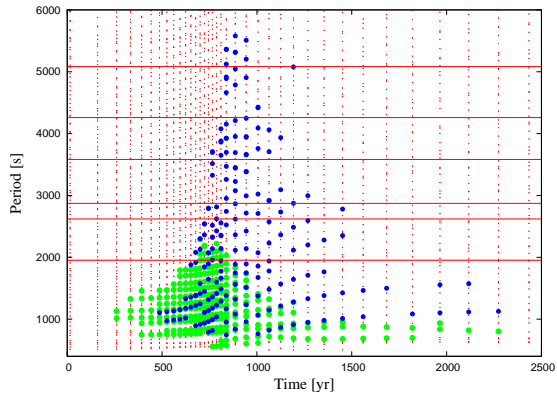}
\caption{Effect of shifting the location of the burning shell on the
  periods of unstable modes in the first subflash. Blue points
  indicate modes destabilized when the burning shell is artificially
  shifted to $\log q\sim -0.37$. Green points show the unstable modes
  in our standard sequence (burning shell at $\log q\sim -0.21$; see
  Fig. \ref{Fig:modos}). Lines indicate the periods detected in LS
  IV-14$^\circ$116, with their width being propotional to the
  amplitude of each mode \citep{2011ApJ...734...59G}.}
\label{Fig:experimento} 
\end{center}
\end{figure}
As mentioned in previous paragraphs, our results suggest that the
periods of unstable modes become shorter when the star experiences
deeper subflashes, that is, when the burning shells move towards the
core. One might then wonder where should the shell be located in order
to excite modes with periods embracing the whole range of periods
observed in LS IV-14$^\circ$116.  In this connection, we performed a
numerical experiment by artificially shifting outwards the location of
the burning shell in our nonadiabatic computations. As shown in
Fig. \ref{Fig:experimento}, it would be enough to shift the location
of the shell to $\log q=\sim -0.37$ ($q:=1-m(r)/M_\star$) to excite
modes with periods covering the complete range of observed periods of
LS IV-14$^\circ$116. Interestingly enough, this location is completely
within the range of locations of the burning shells in the merger
scenario (see figures 2 and 3 of
\citealt{2000MNRAS.313..671S}). Whether this means that LS
IV-14$^\circ$116 is a post-merger object or that He-shell subflashes
develop differently from that current stellar evolution models predict
(as suggested by recent 3D hydro-simulations of the event;
\citealt{2009A&A...501..659M}) should be object of further study.

\section{Conclusions}
We have shown that the $\epsilon$-mechanism is able to excite
pulsations in the He-sdB star LS IV-14$^\circ$116. Although our
computations have been performed within the hot-flasher scenario, the
existence of He-shell subflashes in both the hot-flasher and merger
scenarios makes our conclusions valid regardless the actual origin of
LS IV-14$^\circ$116.

Our results strongly suggest that long period $g$-mode pulsations
exhibited by LS IV-14$^\circ$116 are driven by the
$\epsilon$-mechanism. If so, it is the first confirmation that the
$\epsilon$-mechanism can drive pulsations in stars. Interestingly,
if LS IV-14$^\circ$116 is a star in a fast evolutionary phase then
iron and nickel might not have had enough time to accumulate in the
driving region and drive $p$-mode pulsations, thus explaining also the
intriguing absence of these modes in LS IV-14$^\circ$116 lightcurve.

The fact that our model covers only partially the range of periods
observed in LS IV-14$^\circ$116 might be related to the possibility
that the star is a post-merger object, in which the locations of the
He-flash shells are different from those of hot-flasher objects. In
fact, the numerical experiment performed in the previous section
suggest that nonadiabatic computations of the $\epsilon$-mechanism on
post-merger stellar models might show better agreement with the
observed periods of LS IV-14$^\circ$116. Computations of the
$\epsilon$-mechanism in post-merger He-sdB stellar models would be
very valuable.

In short, we have shown that the LS IV-14$^\circ$116 atmospheric values
($T_{\rm eff}$, $g$ and $n_{\rm He}$), the existence of long period $g$-mode
pulsations, and possibly the absence of the classic $\kappa$-mechanism $p$-mode
pulsations can be simultaneously explained if LS IV-14$^\circ$116 is
as star undergoing a He-shell flash in its way towards the EHB.

\acknowledgments 
 This work has been supported by grants PIP
112-200801-00904 and PICT-2010-0861 from CONICET and ANCyT,
respectively. 
M3B 
thanks both the
Varsavsky Foundation and the organizers of the Fifth Meeting on Hot
Subdwarf Stars \& Related Objects for the financial assistance that
allowed him to attend the meeting, where the central idea of this
article was conceived. 
This research has made use of NASA's Astrophysics Data System.


\begin{thebibliography}{22}
\expandafter\ifx\csname natexlab\endcsname\relax\def\natexlab#1{#1}\fi

\bibitem[{{Ahmad} \& {Jeffery}(2005)}]{2005A&A...437L..51A}
{Ahmad}, A., \& {Jeffery}, C.~S. 2005, \aap, 437, L51

\bibitem[{{Ahmad} {et~al.}(2004){Ahmad}, {Jeffery}, {Solheim}, \&
  {{\O}stensen}}]{2004Ap&SS.291..435A}
{Ahmad}, A., {Jeffery}, C.~S., {Solheim}, J.-E., \& {{\O}stensen}, R. 2004,
  \apss, 291, 435

\bibitem[{{Brown} {et~al.}(2001){Brown}, {Sweigart}, {Lanz}, {Landsman}, \&
  {Hubeny}}]{2001ApJ...562..368B}
{Brown}, T.~M., {Sweigart}, A.~V., {Lanz}, T., {Landsman}, W.~B., \& {Hubeny},
  I. 2001, \apj, 562, 368

\bibitem[{{Charpinet} {et~al.}(1997){Charpinet}, {Fontaine}, {Brassard},
  {Chayer}, {Rogers}, {Iglesias}, \& {Dorman}}]{1997ApJ...483L.123C}
{Charpinet}, S., {Fontaine}, G., {Brassard}, P., {Chayer}, P., {Rogers}, F.~J.,
  {Iglesias}, C.~A., \& {Dorman}, B. 1997, \apjl, 483, L123+

\bibitem[{{C{\'o}rsico} {et~al.}(2006){C{\'o}rsico}, {Althaus}, \& {Miller
  Bertolami}}]{2006A&A...458..259C}
{C{\'o}rsico}, A.~H., {Althaus}, L.~G., \& {Miller Bertolami}, M.~M. 2006,
  \aap, 458, 259

\bibitem[{{C{\'o}rsico} {et~al.}(2009){C{\'o}rsico}, {Althaus}, {Miller
  Bertolami}, {Gonz{\'a}lez P{\'e}rez}, \& {Kepler}}]{2009ApJ...701.1008C}
{C{\'o}rsico}, A.~H., {Althaus}, L.~G., {Miller Bertolami}, M.~M.,
  {Gonz{\'a}lez P{\'e}rez}, J.~M., \& {Kepler}, S.~O. 2009, \apj, 701, 1008

\bibitem[{{Fontaine} {et~al.}(2003){Fontaine}, {Brassard}, {Charpinet},
  {Green}, {Chayer}, {Bill{\`e}res}, \& {Randall}}]{2003ApJ...597..518F}
{Fontaine}, G., {Brassard}, P., {Charpinet}, S., {Green}, E.~M., {Chayer}, P.,
  {Bill{\`e}res}, M., \& {Randall}, S.~K. 2003, \apj, 597, 518

\bibitem[{{Fontaine} {et~al.}(2008){Fontaine}, {Brassard}, {Charpinet},
  {Green}, {Chayer}, {Randall}, \& {van Grootel}}]{2008ASPC..392..231F}
{Fontaine}, G., {Brassard}, P., {Charpinet}, S., {Green}, E.~M., {Chayer}, P.,
  {Randall}, S.~K., \& {van Grootel}, V. 2008, in Astronomical Society of the
  Pacific Conference Series, Vol. 392, Hot Subdwarf Stars and Related Objects,
  ed. {U.~Heber, C.~S.~Jeffery, \& R.~Napiwotzki}, 231--+

\bibitem[{{Green} {et~al.}(2003){Green}, {Fontaine}, {Reed}, {Callerame},
  {Seitenzahl}, {White}, {Hyde}, {{\O}stensen}, {Cordes}, {Brassard}, {Falter},
  {Jeffery}, {Dreizler}, {Schuh}, {Giovanni}, {Edelmann}, {Rigby}, \&
  {Bronowska}}]{2003ApJ...583L..31G}
{Green}, E.~M., {et~al.} 2003, \apjl, 583, L31

\bibitem[{{Green} {et~al.}(2011){Green}, {Guvenen}, {O'Malley}, {O'Connell},
  {Baringer}, {Villareal}, {Carleton}, {Fontaine}, {Brassard}, \&
  {Charpinet}}]{2011ApJ...734...59G}
---. 2011, \apj, 734, 59

\bibitem[{{Kawaler} {et~al.}(1986){Kawaler}, {Winget}, {Hansen}, \&
  {Iben}}]{1986ApJ...306L..41K}
{Kawaler}, S.~D., {Winget}, D.~E., {Hansen}, C.~J., \& {Iben}, Jr., I. 1986,
  \apjl, 306, L41

\bibitem[{{Kilkenny} {et~al.}(2010){Kilkenny}, {Fontaine}, {Green}, \&
  {Schuh}}]{2010IBVS.5927....1K}
{Kilkenny}, D., {Fontaine}, G., {Green}, E.~M., \& {Schuh}, S. 2010,
  Information Bulletin on Variable Stars, 5927, 1

\bibitem[{{Kilkenny} {et~al.}(1997){Kilkenny}, {Koen}, {O'Donoghue}, \&
  {Stobie}}]{1997MNRAS.285..640K}
{Kilkenny}, D., {Koen}, C., {O'Donoghue}, D., \& {Stobie}, R.~S. 1997, \mnras,
  285, 640

\bibitem[{{Lanz} {et~al.}(2004){Lanz}, {Brown}, {Sweigart}, {Hubeny}, \&
  {Landsman}}]{2004ApJ...602..342L}
{Lanz}, T., {Brown}, T.~M., {Sweigart}, A.~V., {Hubeny}, I., \& {Landsman},
  W.~B. 2004, \apj, 602, 342

\bibitem[{{Lenain} {et~al.}(2006){Lenain}, {Scuflaire}, {Dupret}, \&
  {Noels}}]{2006CoAst.147...93L}
{Lenain}, G., {Scuflaire}, R., {Dupret}, M.-A., \& {Noels}, A. 2006,
  Communications in Asteroseismology, 147, 93

\bibitem[{{Miller Bertolami} {et~al.}(2008){Miller Bertolami}, {Althaus},
  {Unglaub}, \& {Weiss}}]{2008A&A...491..253M}
{Miller Bertolami}, M.~M., {Althaus}, L.~G., {Unglaub}, K., \& {Weiss}, A.
  2008, \aap, 491, 253

\bibitem[{{Moc{\'a}k} {et~al.}(2009){Moc{\'a}k}, {M{\"u}ller}, {Weiss}, \&
  {Kifonidis}}]{2009A&A...501..659M}
{Moc{\'a}k}, M., {M{\"u}ller}, E., {Weiss}, A., \& {Kifonidis}, K. 2009, \aap,
  501, 659

\bibitem[{{Naslim} {et~al.}(2010){Naslim}, {Jeffery}, {Ahmad}, {Behara}, \&
  {{\c S}ah{\`i}n}}]{2010MNRAS.409..582N}
{Naslim}, N., {Jeffery}, C.~S., {Ahmad}, A., {Behara}, N.~T., \& {{\c
  S}ah{\`i}n}, T. 2010, \mnras, 409, 582

\bibitem[{{Naslim} {et~al.}(2011){Naslim}, {Jeffery}, {Behara}, \&
  {Hibbert}}]{2011MNRAS.412..363N}
{Naslim}, N., {Jeffery}, C.~S., {Behara}, N.~T., \& {Hibbert}, A. 2011, \mnras,
  412, 363

\bibitem[{{Pesnell}(1987)}]{1987ApJ...314..598P}
{Pesnell}, W.~D. 1987, \apj, 314, 598

\bibitem[{{Saio} \& {Jeffery}(2000)}]{2000MNRAS.313..671S}
{Saio}, H., \& {Jeffery}, C.~S. 2000, \mnras, 313, 671

\bibitem[{{Unno} {et~al.}(1989){Unno}, {Osaki}, {Ando}, {Saio}, \&
  {Shibahashi}}]{1989nos..book.....U}
{Unno}, W., {Osaki}, Y., {Ando}, H., {Saio}, H., \& {Shibahashi}, H. 1989,
  {Nonradial oscillations of stars}, ed. {Unno, W., Osaki, Y., Ando, H., Saio,
  H., \& Shibahashi, H.}

\end{thebibliography}

\end{document}